\PassOptionsToPackage{unicode}{hyperref}
\PassOptionsToPackage{hyphens}{url}
\PassOptionsToPackage{dvipsnames,svgnames,x11names}{xcolor}
\documentclass[
  12pt]{article}
\usepackage{amsmath,amssymb}
\usepackage{iftex}
\ifPDFTeX
  \usepackage[T1]{fontenc}
  \usepackage[utf8]{inputenc}
  \usepackage{textcomp} 
\else 
  \usepackage{unicode-math}
  \defaultfontfeatures{Scale=MatchLowercase}
  \defaultfontfeatures[\rmfamily]{Ligatures=TeX,Scale=1}
\fi
\usepackage{lmodern}
\ifPDFTeX\else  
\fi
\IfFileExists{upquote.sty}{\usepackage{upquote}}{}
\IfFileExists{microtype.sty}{
  \usepackage[]{microtype}
  \UseMicrotypeSet[protrusion]{basicmath} 
}{}
\makeatletter
\@ifundefined{KOMAClassName}{
  \IfFileExists{parskip.sty}{%
    \usepackage{parskip}
  }{
    \setlength{\parindent}{0pt}
    \setlength{\parskip}{6pt plus 2pt minus 1pt}}
}{
  \KOMAoptions{parskip=half}}
\makeatother
\usepackage{xcolor}
\makeatletter
\ifx\paragraph\undefined\else
  \let\oldparagraph\paragraph
  \renewcommand{\paragraph}{
    \@ifstar
      \xxxParagraphStar
      \xxxParagraphNoStar
  }
  \newcommand{\xxxParagraphStar}[1]{\oldparagraph*{#1}\mbox{}}
  \newcommand{\xxxParagraphNoStar}[1]{\oldparagraph{#1}\mbox{}}
\fi
\ifx\subparagraph\undefined\else
  \let\oldsubparagraph\subparagraph
  \renewcommand{\subparagraph}{
    \@ifstar
      \xxxSubParagraphStar
      \xxxSubParagraphNoStar
  }
  \newcommand{\xxxSubParagraphStar}[1]{\oldsubparagraph*{#1}\mbox{}}
  \newcommand{\xxxSubParagraphNoStar}[1]{\oldsubparagraph{#1}\mbox{}}
\fi
\makeatother

\usepackage{longtable,booktabs,array}
\usepackage{calc} 
\usepackage{etoolbox}
\makeatletter
\patchcmd\longtable{\par}{\if@noskipsec\mbox{}\fi\par}{}{}
\makeatother
\IfFileExists{footnotehyper.sty}{\usepackage{footnotehyper}}{\usepackage{footnote}}
\makesavenoteenv{longtable}
\usepackage{graphicx}
\makeatletter
\def\maxwidth{\ifdim\Gin@nat@width>\linewidth\linewidth\else\Gin@nat@width\fi}
\def\maxheight{\ifdim\Gin@nat@height>\textheight\textheight\else\Gin@nat@height\fi}
\makeatother
\setkeys{Gin}{width=\maxwidth,height=\maxheight,keepaspectratio}
\makeatletter
\def\fps@figure{htbp}
\makeatother

\makeatletter
\@ifpackageloaded{caption}{}{\usepackage{caption}}
\AtBeginDocument{%
\ifdefined\contentsname
  \renewcommand*\contentsname{Table of contents}
\else
  \newcommand\contentsname{Table of contents}
\fi
\ifdefined\listfigurename
  \renewcommand*\listfigurename{List of Figures}
\else
  \newcommand\listfigurename{List of Figures}
\fi
\ifdefined\listtablename
  \renewcommand*\listtablename{List of Tables}
\else
  \newcommand\listtablename{List of Tables}
\fi
\ifdefined\figurename
  \renewcommand*\figurename{Figure}
\else
  \newcommand\figurename{Figure}
\fi
\ifdefined\tablename
  \renewcommand*\tablename{Table}
\else
  \newcommand\tablename{Table}
\fi
}
\@ifpackageloaded{float}{}{\usepackage{float}}
\floatstyle{ruled}
\@ifundefined{c@chapter}{\newfloat{codelisting}{h}{lop}}{\newfloat{codelisting}{h}{lop}[chapter]}
\floatname{codelisting}{Listing}

\makeatother
\makeatletter
\@ifpackageloaded{caption}{}{\usepackage{caption}}
\@ifpackageloaded{subcaption}{}{\usepackage{subcaption}}
\makeatother

\ifLuaTeX
  \usepackage{selnolig}  
\fi
\usepackage[]{natbib}
\usepackage{bookmark}

\IfFileExists{xurl.sty}{\usepackage{xurl}}{} 
\hypersetup{
  pdftitle={Title},
  pdfauthor={Author 1; Author 2},
  pdfkeywords={3 to 6 keywords, that do not appear in the title},
  colorlinks=true,
  linkcolor={blue},
  filecolor={Maroon},
  citecolor={Blue},
  urlcolor={Blue},
  pdfcreator={LaTeX via pandoc}}

\newcommand{\anon}{1}

\newcommand{\methodname}[1]{\mbox{ARGEN}}

\usepackage{enumerate}
\usepackage[font=tiny,labelfont=bf]{caption}
\usepackage{tikz}
\usepackage{tikz-network}
\usepackage{pgfplots}
\usepackage{graphicx}
\usepackage{amsmath}
\usepackage{amsbsy}
\usepackage{amssymb}
\usepackage{amsfonts}
\usepackage{mathtools}
\usepackage{booktabs}
 \usepackage{relsize}
\usepackage{array}
\usepackage{comment}
\usepackage{bm}
\usepackage{algpseudocode}

\usepackage[ruled,vlined]{algorithm2e}
\algnewcommand\INPUT{\item[\textbf{Input:}]}%
\algnewcommand\OUTPUT{\item[\textbf{Output:}]}%
\usepackage{xr}
\newcommand{\ind}{\perp\!\!\!\!\perp} 
 
\usepackage{cleveref}
\usepgfplotslibrary{groupplots,dateplot}
\usetikzlibrary{patterns,shapes.arrows}
\usetikzlibrary{arrows.meta}
\pgfplotsset{compat=1.17}
\usepackage{amsthm}

\def\D{{\mathbf D}}

\def\0{{\mathbf 0}}

\definecolor{cerulean}{rgb}{0.0, 0.48, 0.65}
\newtheorem{thm}{Theorem}

\newtheorem{assumption}{Assumption}

\newtheorem{rmk}{Remark}

\crefname{figure}{}{}
\crefname{algorithm}{}{}
\Crefname{algorithm}{}{}
\Crefname{figure}{}{}
\Crefname{section}{}{}

\newcommand{\ignore}[1]{}

\begin{document}

\def\spacingset#1{\renewcommand{\baselinestretch}%
{#1}\small\normalsize} \spacingset{1}



\if1\anon
{
  \title{\bf 
  Confounder-robust causal discovery and inference in Perturb-seq using proxy and instrumental variables}
  \author{Kwangmoon Park \hspace{.2cm}
    and \hspace{.2cm} Hongzhe Li \thanks{
    Corresponding author}\\
    \hspace{.2cm}
    Department of Biostatistics, Epidemiology, and Informatics,\\ University of Pennsylvania, Philadelphia, PA, USA, 19104
    }
  \maketitle
} \fi

\if0\anon
{
  \bigskip
  \bigskip
  \bigskip
  \begin{center}
    {\LARGE\bf Confounder-robust causal discovery and inference in Perturb-seq using proxy and instrumental variables}
\end{center}
  \medskip
} \fi

\bigskip
\begin{abstract}
Emerging single-cell technologies that combine CRISPR-based genetic perturbations with single-cell RNA sequencing, such as Perturb-seq, offer unprecedented opportunities to uncover cause-and-effect relationships among genes. Nonetheless, Perturb-seq experiments are subject to unobserved factors that, if not properly handled, can severely bias the inferred causal relationships between genes. These latent factors may arise not only from intrinsic molecular features of the regulatory elements, but also from unmeasured genes omitted due to cost-constrained experimental designs. Although methods for analyzing large-scale Perturb-seq data are rapidly maturing, approaches that explicitly account for such unobserved confounders when inferring causal gene networks are still lacking. Here, we propose a novel approach to accurately reconstruct causal gene networks from Perturb-seq data even when important confounders are missing. Our framework leverages proxy and instrumental variable strategies to exploit the rich information embedded in the perturbations, enabling unbiased estimation of the underlying directed acyclic graph (DAG) of gene expression. Applications to both comprehensive synthetic data and real CRISPR interference experiments in K562 cells demonstrate that our method outperforms baseline approaches that lack principled adjustments for unmeasured confounding, yielding more accurate and biologically relevant recovery of the true causal DAGs.
\end{abstract}
\noindent%
\vfill

\newpage
\spacingset{1.8} 

\section{Introduction}

A central objective in modern genomics is to recover the directed structure of gene–gene regulation that governs cellular function. Single-cell CRISPR perturbation assays coupled with scRNA-seq, popularized by Perturb-seq, deliver interventional readouts at single-cell resolution across thousands of targeted knockdowns or activations in pooled format \citep{dixit2016perturb, replogle2022mapping}. Beyond revealing heterogeneous and context-specific responses at the single cell levels, these technologies create large-scale datasets that, in principle, permit causal discovery of causal gene networks (CGNs) rather than mere association maps. Such CGNs are often presented as directed acyclic graphs (DAGs), where nodes are genes and directed edges present regulatory relationships. 

From a statistical perspective, learning DAGs from observational has been studied widely in literature. Classical constraint-based procedures such as the PC algorithm \citep{spirtes2000causation} rely on conditional independence tests to identify Markov equivalence classes under assumptions of causal sufficiency and faithfulness. Score-based approaches such as the Greedy Equivalence Search (GES) \citep{chickering2002Optimal} optimize penalized likelihoods over equivalence classes, while hybrid approaches combine constraint and score elements. Continuous optimization frameworks such as NOTEARS \citep{zheng2018dags} recasts DAG learning as smooth optimization problems, enabling scalable inference in high dimensions. While these frameworks have advanced causal discovery in observational data, they generally assume that all confounders are observed.
When interventional data are available, causal discovery can, in principle, be strenghtened. Algorithms such as GIES \citep{hauser2012characterization} extend GES to integrate intervention targets for orienting edges and improving identifiability.

However, these methods typically presume that all participating nodes of the underlying DAG are observed in the data, without omitted factors that can alter the distributions of confounded nodes. In pooled CRISPR single-cell screens, this assumption is likely to be violated (Section~\ref{sec:motivating_data} for details).  Ignoring unobserved factors may result in biased inference of causal relationships in DAGs. 
In addition, technical covariates such as sequencing depth, capture efficiency, and batch influence both gene expression and gRNA detection. 

Several computational tools have been proposed to model Perturb-seq data. The original Perturb-seq studies introduced MIMOSCA for regularized linear modeling of perturbation effects \citep{dixit2016perturb}. Subsequent methods such as scMAGeCK \citep{yang2020scmageck} extended genotype-to-multivariate-phenotype mapping for single-cell CRISPR screens. More recently, SCEPTRE \citep{barry2021sceptre} provided a statistically calibrated approach using conditional resampling to correct for confounding from technical covariates, substantially improving type I error control and sensitivity in detercting the affected genes from perturbation. Despite these advances, current frameworks primarily focus on estimating pairwise perturbation–response effects and do not naturally extend to reconstructing global regulatory DAGs. Moreover, robust computational methods for estimating DAGs in the presence of unobserved factors are still lacking.

In this paper, we propose \methodname{} for \textbf{A}rbitrary-confounder \textbf{R}obust causal \textbf{GE}ne \textbf{N}etwork, as a statistical framework for recovering causal gene network from Perturb-seq experiments that is robust to unobserved confounders.  \methodname{} explicitly leverages the interventional design of CRISPR screens, and 
 adopts an instrumental variable strategy that uses proxy gene expression to identify causal gene networks. Guide detections serve as instrumental variables, and proxies for hidden gene expression provide robustness against confounding.

We extend the classical measurement model for scRNA-seq data \citep{huang2018saver,sarkar2021separating} to a setting that explicitly reflects both the underlying gene regulatory network and the perturbation structure inherent to Perturb-seq data. Specifically, we employ a structural equation model (SEM) with embedded perturbation terms to represent the causal relationships among gene expressions under CRISPR interventions.
Under the assumed perturbation model, we investigate conditions for identification of the ancestors and descendants of each perturbed (intervened) target gene, as well as the identification of the causal ordering and the underlying DAG. We further develop novel algorithm for estimating descendant sets and parental relationships, with provable guarantees that the resulting estimated structure yields a valid DAG, even in finite-sample settings.

Through extensive simulation studies and applications to real Perturb-seq datasets on K562 cell lines, we show that our approach consistently improves accuracy in recovering the true causal structure compared to existing methods that ignore unmeasured confounding. Our results highlight the importance of causal-robust statistical modeling in translating single-cell CRISPR screens data  into reliable and interpretable gene causal networks.

\section{A statistical framework for Perturb-seq data analysis}

\begin{figure}[!ht]\centering
   \includegraphics[width=1\textwidth]{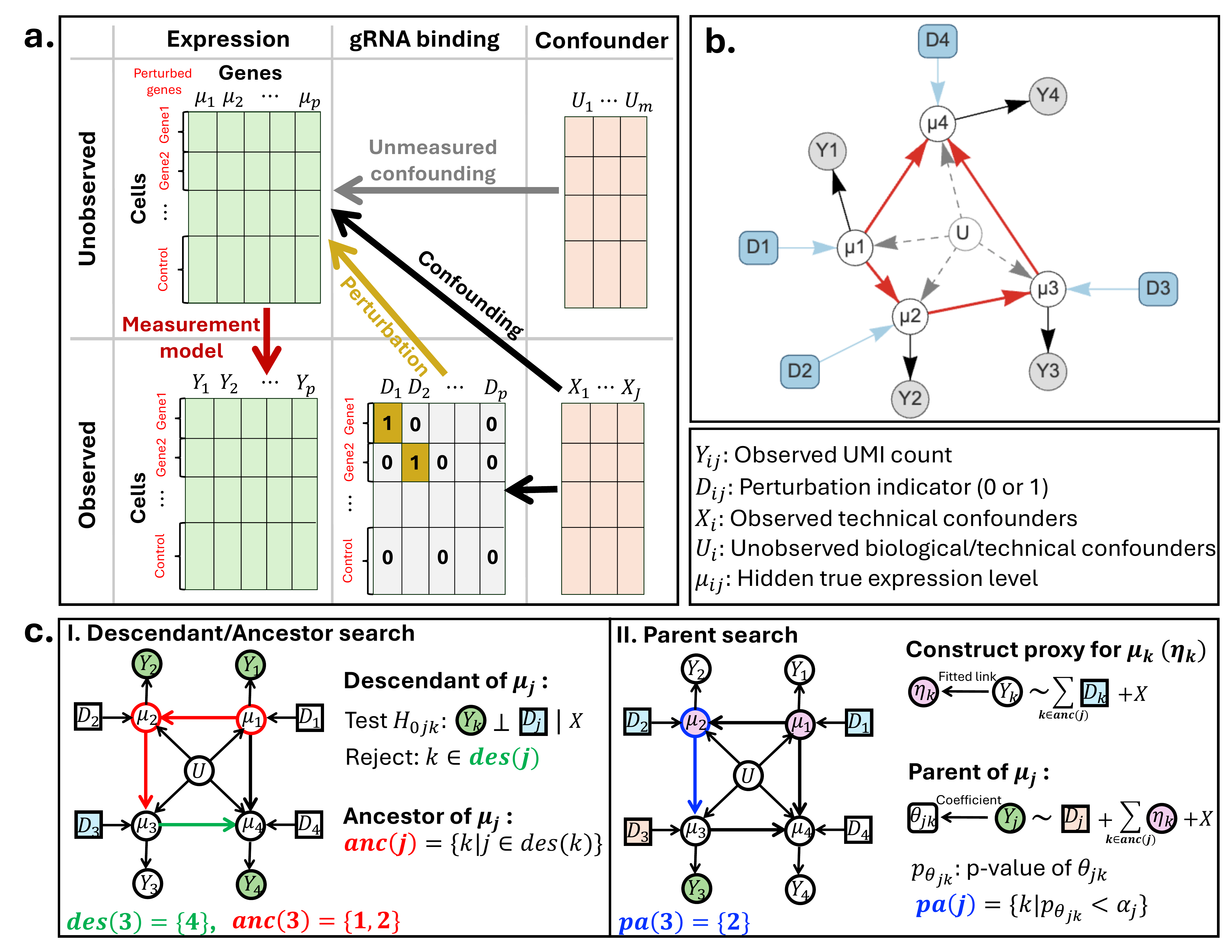}
\caption{\textbf{Schematic overview of \methodname{}.}
\textbf{a.} \methodname{} takes as input single-cell--level UMI counts
$Y=(Y_1,\ldots,Y_p)\in\mathbb{R}^p$, gRNA binding indicators
$D=(D_1,\ldots,D_p)\in\{0,1\}^p$, and technical confounders
$X=(X_1,\ldots,X_J)\in\mathbb{R}^J$, all of which are observable.
Each cell receives a perturbation on a single gene.
\methodname{} relies on a classical measurement model
\citep{sarkar2021separating} that generates UMI counts
$(Y_1,\ldots,Y_p)$ from latent true expression levels
$(\mu_1,\ldots,\mu_p)$.
Each $\mu_j$ directly receives the perturbation treatment and is
influenced by both observed ($X$) and unobserved confounders
$(U=(U_1,\ldots,U_m))$. The gRNA binding indicator is also associated with the observed technical factors $X$. \textbf{b.} A simplified DAG with $p=4$, representing the structural equation model of \methodname{}. The goal of \methodname{} is to learn the (red) edges among the unobserved true expression variables. For simplicity, $X$ is omitted from the visualization. 
\textbf{c.} Overview of \methodname{}'s DAG search procedure. Left: \methodname{} first estimates the descendant sets of each $\mu_j$ by testing~\eqref{condind_null} across $k\neq j$. The estimated $\{\mathrm{des}(j)\mid j\in[p]\}$ are then passed to Algorithm~\ref{supp-alg:DAG_search}. Right: \methodname{} subsequently identifies the parent set $pa(j)$ for each $j\in[p]$, proceeding from the bottom layer to the top within Algorithm~\ref{supp-alg:DAG_search}. Parents are estimated via the following steps:
(i) construct proxy measurements $\eta_k$ for the true expression $\mu_k$ for all $k\in anc(j)$; (ii) solve a QMLE to estimate $\theta_{jk}$ for each gene $j$; and (iii) adjust the p-values of $\theta_{jk}$, denoted by $p_{\theta_{jk}}$, using Online FDR~\citep{zrnic2020power} across $j\in[p]$ to obtain the final estimate of $pa(j)$.
} \label{fig:main_summary}
\end{figure}

\subsection{Motivation: Perturb-seq data and causal gene network}\label{sec:motivating_data}
This paper is motivated by the problem of learning causal gene networks among sets of essential genes based on Perturb-seq data \citep{replogle2022mapping}. The data was generated by treating K562 cells with CRISPR interference (CRISPRi) to systematically knock down essential genes one at a time, allowing for the measurement of subsequent expression changes across the transcriptome at the single cell resolution. The targeted essential genes are involved in a wide array of fundamental cellular processes, most notably transcription and chromatin regulation, DNA replication, translation initiation, RNA processing, and ribosome biogenesis. After standard quality control, the dataset includes 1,869 perturbed genes with their transcriptomic profiles. The number of cells per perturbation ranges from 5 to 1,996, while the control group with non-targeting guides has 10,691 cells. Unique molecular identifier (UMI) counts per gene range from 0 to 1,114. Such a large-scale perturbation dataset provides a unique opportunity to infer causal regulatory networks by leveraging intervention-induced changes in gene expression at single-cell resolution.

Solving this problem holds profound practical and clinical significance, since essential genes govern the fundamental processes required for cellular survival, proliferation, and defense response. Moving beyond correlational co-expression to accurately recover causal regulatory architecture is vital for understanding how genetic perturbations cascade through networks to drive complex traits, which is critical for identifying true therapeutic targets.

The problem of elucidating the causal structure of genes based on this interventional data can be formulated as estimating a DAG among the underlying true gene expression levels \citep{huang2018saver,sarkar2021separating}. While intuitive, the data presents severe practical challenges: (i) biological heterogeneity, such as cell cycle phase, chromatin accessibility, and lineage differences, introduces latent variation influencing cellular transcriptomic profiles; and (ii) due to experimental cost considerations, CRISPR screens typically target key genes of interest, yielding incomplete datasets with omitted variables.

We overcome these challenges by translating these data-driven problems into the statistical framework outlined in the following sections. By doing so, we explicitly address the core scientific questions:
(a) What is the causal architecture of K562 essential processes, and can we systematically isolate true causal relationships from spurious associations? (Sections~\ref{sec:direction_robustness} and \ref{sec:structure_inter})
(b) Can we recover a robust causal gene network despite the limited set of variables captured in Perturb-seq experiments? (Section~\ref{sec:direction_robustness})
(c) What are the epigenetic and genomic characteristics of the causal regulatory relationships learned from this interventional transcriptomic data? (Sections~\ref{sec:epi_3d} and \ref{sec:func_epi_eval})

\subsection{Data observed in perturb-seq experiments and DAG}\label{subsection:notation}

We now provide a statistical framework to address the key challenges stemming from our motivating dataset. We first translate the data observed from Perturb-seq experiments into statistical objects, and then introduce a statistical model in the following section. 

We denote the UMI count of $p$ observed genes of a cell in a Perturb-seq data as $Y_{1},\cdots,Y_{p}$. We also denote guide RNA (gRNA) perturbation indicator $D_j\in \{0,1\}$, where $j\in\{1\cdots,p\}$, and  $D_j=1$ indicating that the cell is induced by a guide that perturbs the gene $Y_j$ $\forall j\in[p]$.  We further assume $\sum_{j=1}^pD_j\in\{0,1\}$, meaning that we target a single gene at a time or do not target any genes. A control cell with $D_1=D_2\cdots=D_p=0$ indicates that the cell is induced by a non-targeting guide that perturbs no genes. We define the unobserved biological confounders such as accessibility level of enhancer regions, hidden sub cell types or even unmeasured expression level of other confounding genes as $U=(U_1,\cdots,U_m)^T$ that influences transcriptomic profile of 
the $p$ genes. Lastly, we  define the observable cell-specific covariates as $X=(X_1,\cdots,X_J)^T$, including  technical factors that could influence both the expression level and $D_j$ (\textbf{Fig.}~\ref{fig:main_summary}a). 

We use  DAG  to represent the causal network among the expression levels of $p$ genes. Specifically, a DAG $G$ is a pair $(V, E)$ consisting of a finite set of nodes $V (=V(G))= \{1, \ldots, p\}$ and a set of directed edges $E \subseteq V \times V$ such that the graph contains no directed cycles. That is, there does not exist a sequence of distinct nodes $i_1, i_2, \ldots, i_k$ with directed edges $i_1 \rightarrow i_2 \rightarrow \cdots \rightarrow i_k \rightarrow i_1.$ Each node $j \in V$ typically represents a random variable $\mu_j$ that represents the true expression of gene $j$ in a given cell, and a directed edge $i \rightarrow j$ encodes a potential direct causal or regulatory influence of $\mu_i$ on $\mu_j$.
The acyclicity property implies that the variables can be topologically ordered so that every variable depends only on its predecessors in that order. This property allows one to represent the joint distribution of $\mu_j$s via a structural equation model (SEM) of the form
\[
\mu_j = f_j(\mu_{\mathrm{pa}(j)}, \varepsilon_j), \quad j = 1, \ldots, p,
\]
where $\mathrm{pa}(j)$ denotes the set of parents of node $j$, $f_j(\cdot)$ are measurable functions, and the noise variables $\varepsilon_j$ are mutually independent.

We denote a set of sequential numbers, e.g., $\{1,\cdots,p\},$ as $[p]$, and use $V$, $V(G)$, or $[p]$ interchangeably. We also denote $anc(j)\subset [p]$ and $des(j)\subset [p]$ as the ancestors and descendants of a node $j$, respectively. Ancestors are nodes that have a directed path to $j$, and descendants are nodes reachable from $j$.

In the following section, we introduce a statistical model for Perturb-seq data analysis. 

\subsection{Classical measurement model for scRNA-seq and its extension to Perturb-seq data}\label{subsection:sem}

We  consider the standard and commonly applied measurement model 
for scRNA-seq count data \citep{huang2018saver,sarkar2021separating}. Specifically, we define the hidden expression level of a gene $j\in [p]$ in a cell $i\in[N]$ as $\lambda_{ij}$ and consider the observed UMI count of the gene in the cell as a random Poisson draw from the distribution 
\begin{align}\label{target_gene_model}
Y_{ij}|\lambda_{ij},\ell_i&\stackrel{ind}{\sim} \mbox{Poisson}\left(\ell_i\lambda_{ij}\right),
\end{align}
where $\ell_i$ denotes the sequencing depth of a cell $i$, which can be typically estimated by the total UMI count for a cell $i$ \citep{vallejos2017normalizing} or by alternative approaches such as scran \citep{l2016pooling}. We can consider different types of (prior) expression models for $\lambda_{ij}$ \citep{sarkar2021separating}, such as 
\begin{align*}
\lambda_{ij}&\sim \delta_{\mu_{ij}} \mbox{ (point mass at }\mu_{ij}) \quad \mbox{ or }\\
\lambda_{ij}&\sim \Gamma\left(\mbox{shape}=\phi_j^{-1},\mbox{rate}=\mu_{ij}^{-1}\phi_j^{-1}\right),
\end{align*}
where $\Gamma$ refers to a Gamma distribution. When $\lambda_{ij}$ is integrated out, the former choice gives the observational model $\mbox{Poisson}\left(\ell_i\mu_{ij}\right)$, while the latter gives $\mbox{NB}\left(\mbox{mean}=\ell_i\mu_{ij},\mbox{dispersion}=\phi_j\right)$, where $\mu_{ij}$ represents \textit{true expression level} and $\phi_j$ represents additional overdispersion parameter.

We extend this standard model to the setting that reflects a potential causal gene network and CRISPR perturbation of Perturb-seq data. We specifically employ the expression model with a structure equation model (SEM) specified by a DAG  $G$ with perturbation as follows:
\begin{equation}\label{SEM}
\log{\mu_{ij}}=
\theta_{j0}+\sum_{k\in pa(j)}\log{\mu_{ik}}\theta_{jk}+\tau_j D_{ij}+\beta_j^\top X_i+\gamma_j^\top U_i+\epsilon_{ij}, \mbox{ for } \forall i\in[N], j\in [p].
\end{equation}
where  $\epsilon_{ij}$ is a $i.i.d$ mean zero noise independent of other variables. Our target parameter $\theta_{jk}$ quantifies the influence of gene $k$ to $j$ and $\tau_j$ quantifies how effective the perturbation $D_{ij}$ is for gene $j$. We also note that the binary indicator $D_{ij}$ can depend on the observed technical factors $X_i$ or $l_i$ \citep{barry2024robust}. On the other hand, we assume independency between $D_{ij}$ and $U_i$ given $X_i$ and $l_i$, $(D_{ij}\ind U_i |\ell_i,X_i)$, which implies that the hidden biological factors have less to do with whether or not a gene is perturbed. Lastly, notice that the unobserved confounder $U_i$ can influence $\log{\mu_{ik}}$ for $k\in  pa(j)$ and therefore $\theta_{jk}$ is not in general identifiable. Throughout, we suppress the cell index $i$ when no ambiguity arises. 

The model provides a super-DAG involving $\mu_j,Y_j,X_j,U_j,D_j$ as nodes (\textbf{Figure.}~\ref{fig:main_summary}b). Here, we focus on learning the DAG $G$ among the true expressions $\mu_j$s (red subgraph in \textbf{Figure.}~\ref{fig:main_summary}b). 

With the statistical framework introduced, we aim to address the outlined motivating problems in Section~\ref{sec:motivating_data} by accomplishing the statistical goal of correctly estimating the DAG $G$ and the causal effect $\theta_{jk}$ despite the existence of  unmeasured confounders $U$.

\section{Identifiability of the Causal Network}\label{sec:identify}

Before estimating the DAG among the genes, a DAG identification result must be established under the assumed mdoel and settings. We provide detailed  identification results in Section~\ref{supp-sec:DAG_identification} with several theoretical arguments. One key point is that we  need  the following  much weaker assumption  than the classical \textit{faithfulness} \citep{pearl2009causality} or \textit{influentiality} \citep{tian2013causal}, which turns out to be sufficient for recovering the ancestral information of the entire DAG.

\begin{assumption}[Non-degenerate direct effect]\label{asm:nond}
For any gene $j\in[p]$, we have $\theta_{jk}\neq0$ (non-zero causal effect) for $k\in pa(j)$ and $\tau_j\neq0$ (non-zero intervention effect).
\end{assumption}

 We  show in  Theorem~\ref{thm:des} that the ancestor and descendant nodes of each gene can be identified. This is justified by  introducing a new concept called \textit{exlusive directed path} (Definition~\ref{supp-def:exclusive_directed_path}), which denotes a single unique directed path between two nodes. We then show the existence of such path in the subgraph between any two nodes in Lemma~\ref{supp-lem:exclusive_path} and that influentiality can be achieved in the exclusive directed path in Lemma~\ref{supp-lem:inf_exclusive_path}. 
 
 \begin{thm}[Identifiability of descendant and ancestor nodes]\label{thm:des}
Under Assumption~\ref{asm:nond} and the measurement model with intervention on each node, for all $j\in V(G)$, $des(j)$ and $anc(j)$ are identifiable. 
\end{thm}
 
Given the ancestral information identified, we show that the causal ordering of the DAG and the DAG itself is identifiable given a correct parent set search procedure in Theorem~\ref{supp-thm:order} (Section~\ref{supp-subsec:ordering}). Finally, identification of the parents  $pa(j)$ of each $j$ even with the existence of omitted variables is based on the following key theorem by combining ideas from proxy-variable adjustment and instrumental-variable reasoning  (see  Section~\ref{supp-subsec:proxy} for details). 
\begin{thm}[Identifiability of the parents]\label{thm:parent_identify}
Define $\eta_k(\D,X)\coloneqq \log\mathbb{E}\left[
Y_{k}/\ell|\D,X,\ell
\right]$ as the proxy of $r_k(\D,X)\coloneqq \mathbb{E}[\log{\mu_{k}}|\D,X]$ and $\D\coloneqq (D_1,\cdots, D_p)^T$.
Under   either of the  assumptions $(A1): U=\beta_{0u}+B_u^\top X+\epsilon_u\mbox{ and } \epsilon_u\ind X$ or $(A1^\prime): U\ind X$, for each $j\in [p]$ and for all $k\in anc(j)$, the coefficient $\theta_{jk}$ can be correctly identified through Quasi Maximum Likelihood Estimation (QMLE) problem with the mean function
\begin{equation}\label{glm_qmle}
\log{\mathbb{E}[Y_j|\D,X,\ell]}=\log{\ell}+
\theta_{j0}^*+{\beta^*_j}^\top X+\tau_jD_j+\sum_{k\in pa(j)}\theta_{jk}\eta_k(\D,X),
\end{equation}
equivalently, we can identify $pa(j)$ for all $\forall j\in [p]$, i.e., the DAG can be recovered.

\end{thm}
We provide in Section \ref{sec:parent} on how to estimate the proxies $\eta_k(\D,X)$ from the data and how to identify $pa(j)$ for each node $j$.

\begin{rmk}[Remarks on identifiability assumptions]
Assumption~1 requires a nonzero intervention effect, analogous to the
\emph{relevance} assumption in instrumental variable analysis
\citep{durbin1954errors}. When perturbations are
weak, the downstream effects of $D_j$ on $Y_{des(j)}$ may be difficult to
detect, leading to reduced power for descendant recovery and, consequently,
for parent identification. We therefore recommend preliminary differential
expression analyses to verify that each perturbation induces a measurable
change in its target gene $j\in[p]$.

The assumption $D_j \ind U \mid \ell, X$ is violated when perturbation
assignment is associated with unmeasured factors, for example through cell-state-dependent lentiviral silencing or transcription rates tied to metabolic activity. Such violations may bias both
descendant and parent searches. Although this assumption is generally
untestable, practical diagnostics include examining guide assignment quality and distributions of key cell-level quality-control metrics across perturbations. In our motivating Perturb-seq setting,
direct guide capture  substantially reduces
misassignment and detection errors compared with PolyA expression-based guide
identification approaches, including CROP-seq
\citep{datlinger2017pooled}. 
\end{rmk}

\section{DAG Estimation Based on Perturb-seq Data}

With the identifiability results of a DAG established, we provide specific procedures to estimate the descendants, parents and the final DAG from finite-sample Perturb-seq datasets.

\subsection{Estimation of the descendant sets}\label{sec:est_des}
Results in Section~\ref{supp-subsec:des} show that we can identify the descendants of each gene $j$ by testing for the differential expression of gene $k$ upon perturbing gene $j$:
\begin{equation}\label{condind_null}
H_{0jk}:\mathbb{E}[Y_k|D_j=1,X,\ell,\D_{-j}=\0]=\mathbb{E}[Y_k|D_j=0,X,\ell,\D_{-j}=\0]
\end{equation}
for each $k\neq j$ (\textbf{Fig.}~\ref{fig:main_summary}c). We now provide a way to test this based on the model \eqref{target_gene_model} and \eqref{SEM}.

First, we established in Lemma~\ref{supp-lem_condind} that 
\begin{equation}\label{condind_mean}
\log{\mathbb{E}[Y_k|D_j=d,X,\ell,\D_{-j}=\0]}=\log{\ell}+\theta_{0k}^*+\tau_{jk}^*d+\beta_j^\top X.    
\end{equation}
The result guarantees the validity of QMLE for $Y_k|D_j,X,\ell$ from the cells with gene $j$ being perturbed or for the control cells with $\D_{-j}=\0$. Thus, for testing if the gene $k$ is a descendant of $j$ (\textbf{eqn.}~\eqref{condind_null}), we rely on a GLM, such as Poisson or NB, to estimate the coefficient $\tau_{jk}$ with an appropriate adjustment on the variance. Specifically, we fit 
\begin{equation}
Y_{ik}\stackrel{GLM}{\sim}\mbox{offset}(\log{\ell_i})+1+D_{ij}+X_i,\quad \mbox{ for } i\in \{i\in [N]|D_{ij}=1\}\cup\{i\in [N]|\D_{i}=\0\},
\end{equation}
and obtain the coefficient estimator $\hat{\tau}^*_{jk}$ of $D_{ij}$.

We estimate the variance of $\hat{\tau}^*_{jk}$ using a sandwich estimator. We refer to Section~\ref{supp-sec:sandwich} for details on the variance estimator $\hat{V}_{\tau^*_{jk}\tau^*_{jk}}$ for different choices of the expression models. With the variance estimator, the Wald statistic for testing $H_{0jk}\!:\tau_{jk}^*=0$ can be constructed as
\begin{equation}
z_{\tau^*_{jk}}=\frac{\hat{\tau}_{jk}^*}{\sqrt{\hat{V}_{\tau^*_{jk}\tau^*_{jk}}}},
\end{equation}
which yields asymptotically valid inference even if other parameters are misspecified, 
provided the mean function \eqref{condind_mean} is correctly specified \citep{white1982maximum}. Specifically,  we have
$
z_{\tau^*_{jk}}\stackrel{d}{\rightarrow}N(0,1),
$
as $N\rightarrow \infty$, under $H_{0jk}$. Across across $j,k\in [p]$ for $j\neq k$, we collect the p-values 
\begin{equation}\label{eq:des_p}
p_{jk}\coloneqq 1-\Phi(|z_{\tau^*_{jk}}|)
\end{equation}
and employ \cite{benjamini1995controlling} FDR control at certain pre-specified $\alpha\in(0,1)$.

If we choose to rely on classical influentiality assumption \citep{tian2013causal}, we can finalize the descendant search by defining $\hat{des}(j)\coloneqq \{m\in [p]/\{j\}|
p_{jk}<\alpha_{adj}\}.$ 

Alternatively, if the influentiality assumption is of a concern, we can rely only  on Assumption~\ref{asm:nond} and the identification strategy utilizing `descendants of descendants' introduced in Section~\ref{supp-proof:thm:des}. Specifically, after collecting the intervention-descendants defined as $\hat{des}_I(j)\coloneqq \{m\in [p]/\{j\}|
p_{jk}<\alpha_{adj}\},$ we estimate the descendant set of $j$ as 
\begin{equation*}
\hat{des}(j)=\bigcup_{r \geq 1} \hat{des}_I^{(r)}(j),    
\end{equation*}
where $\hat{des}_I^{(1)}(j) = \hat{des}_I(j)$ and 
$\hat{des}_I^{(r+1)}(j) \;=\; \bigcup_{m \in \hat{des}_I^{(r)}(j)} \hat{des}_I(m)$.

Finally, the ancestor set of $j$, $anc(j)$, is estimated as $\hat{anc}(j)\coloneqq\{
k\in V(G)|j\in\hat{des}(k)
\}$.

\subsection{Estimation of the parent sets}\label{sec:parent}
Given the identified descendants $\hat{des}(j)$ and ancestors $\hat{anc}(j)$ of gene $j$, we next aim to identify the direct parents of the gene $j$ when there exists unmeasured confounder $U$. As shown in Theorem \ref{thm:parent_identify}, 
\ignore{
We  can show (Section~\ref{supp-subsec:proxy} for details) the identifiability of $\theta_{jk}$ from a Quasi Maximum Likelihood Estimation (QMLE) with some regular conditions:
$$Y_j\stackrel{GLM}{\sim}
\mbox{offset}(\log{\ell})+X+D_j+\sum_{k\in pa(j)}r_k(\D,X),
$$ 
if the conditional true expression $r_j(\D,X)\coloneqq \mathbb{E}[\log{\mu_{j}}|\D,X]$ were available.

Nonetheless, $r_j(\D,X)$ is not directly observable, which necessitates considering a surrogate (or proxy) for $r_j(\D,X)$. We therefore established throughout Theorem~\ref{supp-thm:proxy},\ref{supp-thm:parent_identify} that
\begin{equation}\label{glm_qmle}
\log{\mathbb{E}[Y_j|\D,X,\ell]}=\log{\ell}+
\theta_{j0}^*+{\beta^*_j}^\top X+\tau_jD_j+\sum_{k\in pa(j)}\theta_{jk}\eta_k(\D,X),
\end{equation}
for \textit{the population link proxy} of $\log{\mu_k}$, defined as $$\eta_k(\D,X)\coloneqq
\log\mathbb{E}\left[
Y_{k}|\D,X,\ell\right]-\log{\ell},$$ which provided identifiability of $\theta_{jk}$ (Section~\ref{supp-subsec:proxy} for details).}
identification of the parents can be based on the mean function, 
\begin{equation}\label{glm_qmle}
\log{\mathbb{E}[Y_j|\D,X,\ell]}=\log{\ell}+
\theta_{j0}^*+{\beta^*_j}^\top X+\tau_jD_j+\sum_{k\in anc(j)}\theta_{jk}\eta_k(\D,X)
\end{equation}
with $\theta_{jk}=0$ for $k\notin pa(j)$, where  $\eta_k(\D,X)\coloneqq
\log\mathbb{E}\left[
Y_{k}|\D,X,\ell\right]-\log{\ell}$ serves as a proxy of $\log{\mu_k}$. Here, we aim to estimate the parents with $\theta_{jk}\ne 0$  (\textbf{Fig.}~\ref{fig:main_summary}d).

However, in practice, the regressor $\eta_k(\D,X)$ needs to be first predicted from the data before fitting a QMLE in \eqref{glm_qmle}. By \eqref{glm_qmle} and the definition of $\eta_k(\D,X)$, we can derive a recursion
\begin{equation*}
\eta_k(\D,X)=\theta_{k0}+{\beta^*_k}^\top X+\tau_kD_k+\sum_{l\in anc(k)}\theta_{kl}\eta_l(\D,X)
\end{equation*}
with $\theta_{kl}=0$ for $l\notin pa(k)$, which, by structural induction over the DAG, establishes $\eta_k(\D,X)$ as a linear function of $D_k$, $\D_{anc(k)} \coloneqq (D_l)_{l \in \mathrm{anc}(k)} \in \mathbb{R}^{|\mathrm{anc}(k)|}$, and $X$. Specifically, we have
$$\eta_k(\D,X)\coloneqq\log{\mathbb{E}[Y_k|\D,X,\ell]}-\log{\ell}=
(1,D_k,\D_{anc(k)}^\top,X^\top)^\top\xi_k.$$
Thus, before fitting a QMLE for \eqref{glm_qmle}, we first predict the proxy $\eta_k(\D,X)$ as follows
\begin{equation}\label{eq:proxy_est}
\mbox{\textbf{First-stage regression}: }\hat{\eta}_{k}(\D_i,X_i)=(1,D_{ik},\D_{i, 
\hat{anc}(k)}^\top,X_i^\top)^\top\hat{\xi}_k
\end{equation}
with the estimated $\hat{anc}(j)$, where the coefficient $\hat{\xi}_k$ can be estimated through  GLM fit for the QMLE of the mean function, $\log{\mathbb{E}[Y_k|\D,X,\ell]}$, estimation.

We now come back to the original regression problem in \eqref{glm_qmle} 
with the predicted proxy $\hat{\eta}_k(\D,X)$ as the estimated regressors 
\begin{equation}\label{eq:second_stage}
\mbox{\textbf{Second-stage regression}: }Y_{ij}\stackrel{GLM}{\sim} \mbox{offset}(\log{\ell_i})+
X_i+D_{ij}+\sum_{k\in \hat{anc}(j)}\hat{\eta}_k(\D_i,X_i),
\end{equation}

The proposed two-stage framework is closely related in spirit to the classical two-stage least squares approach with valid instrumental variables. Here, $\D_{\mathrm{anc}(j)}$ participates in the entire first-stage regressions across $k$ and serves as a valid instrumental variable for $j$. Specifically, it satisfies: (i) conditional independence from unobserved confounders, $U \ind \D_{anc(j)} \mid X$; (ii) the exclusion restriction, as there is no direct causal path from $\D_{anc(j)}$ to $\mu_j$; and (iii) relevance, since $D_k$ directly perturbs $\mu_{k}$ individually for $k\in anc(j)$.

Here, the uncertainty of $\hat{\eta}_k(\D,X)$ needs to be incorporated for a correct quantification of the standard error of $\hat{\theta}_{jk}$ estimated from QMLE. 
We consider the regression coefficients $\hat{\theta}_{jk}$ for each $\hat{\eta}_k$ based on a GLM, such as Poisson, and quantify the standard error of $\hat{\eta}_k$ based on the \cite{murphy2002estimation} adjustment for estimated regressors (Section~\ref{supp-sec:MT} for details).

Then, under regularity conditions in \cite{murphy2002estimation} for those QMLEs, we have 
\begin{equation}\label{edge_zscore}
z_{\theta_{jk}}=\frac{\hat{\theta}_{jk}}{\sqrt{\hat{V}^{(j)}_{\mathrm{MT},k,k}}}
\xrightarrow{d} N(0,1),    
\end{equation}
as $N\rightarrow \infty$, providing asymptotically valid Wald tests for $\theta_{jk}$ (Fig.~\ref{supp-fig:mt_se}). The corresponding p-values are used in online FDR control \citep{zrnic2020power} in our final DAG search algorithm. The Gaussian approximation for the edge coefficients is suitable for Perturb-seq data, where the number of genes $p$ generally satisfies $p \ll N$. Although we do not adopt this strategy, one could fit a regularized GLM to handle high-dimensional settings.

\subsection{A DAG search algorithm with interventions.}
When the parent information is perfect for all $j\in V(G)$, the full DAG is directly obtainable by concatenating the parent nodes of each node. However, estimation errors in parent identification may occur, and naive concatenation of the estimated parent sets can introduce cycles into the resulting graph. To guarantee that the resulting graph is a valid DAG, we propose a sequential DAG search procedure, detailed in Section~\ref{supp-sec:DAG_search} and Algorithm~\ref{supp-alg:DAG_search}. Briefly, the algorithm leverages the causal ordering information established in Theorem~\ref{supp-thm:order} by constructing the graph starting from sink nodes (nodes with no descendants) and sequentially progressing toward root nodes. 
Furthermore, because the number of tests in the parent search depends on the estimated ancestry set $\hat{anc}(j)$, we employ online batch FDR control \citep{zrnic2020power} to dynamically adjust the significance thresholds across sequentially evaluated response genes. Specifically, based on the computed p-values $p_{\theta_{jk}}\coloneqq1-\Phi(z_{\theta_{jk}})$ and the online-adjusted p-value cut off $\alpha_j$, we define the estimated parent set for gene $j$ as
\begin{equation}\label{eq:parent_fdr}
\hat{pa}(j)\coloneqq \{k\in \hat{anc}(j) \mid p_{\theta_{jk}}<\alpha_j \},
\end{equation}

\section{Simulation}\label{sec:main_simul}
We conducted extensive simulations to evaluate \methodname{}. Specifically, we investigated: \textbf{i)} the robustness of its standard error correction under varying distributional assumptions (Section~\ref{supp-sec:simul_dist_MT}), \textbf{ii)} the accurate recovery of DAG structures and edge coefficients under unmeasured confounding (Section~\ref{supp-sec:simul_DAG}), \textbf{iii)} its robustness and unbiasedness against arbitrarily omitted genes (Section~\ref{supp-sec:simul_omitgenes}), \textbf{iv)} desired $N$ and $p$ ratio (Sections~\ref{supp-sec:simul_extended} and \ref{supp-sec:simul_optimalNandP}) and \textbf{v)} its sensitivity to misspecification in the first stage regression and heterogeneity in the guide efficiency (Sections~\ref{supp-sec:sense_misspecification} and \ref{supp-sec:sense_hetero}). We present the results for \textbf{iii)} here briefly, deferring the details and remaining results to Section~\ref{supp-sec:simul}.

\begin{figure}[!ht]\centering
   \includegraphics[width=1.0\textwidth]{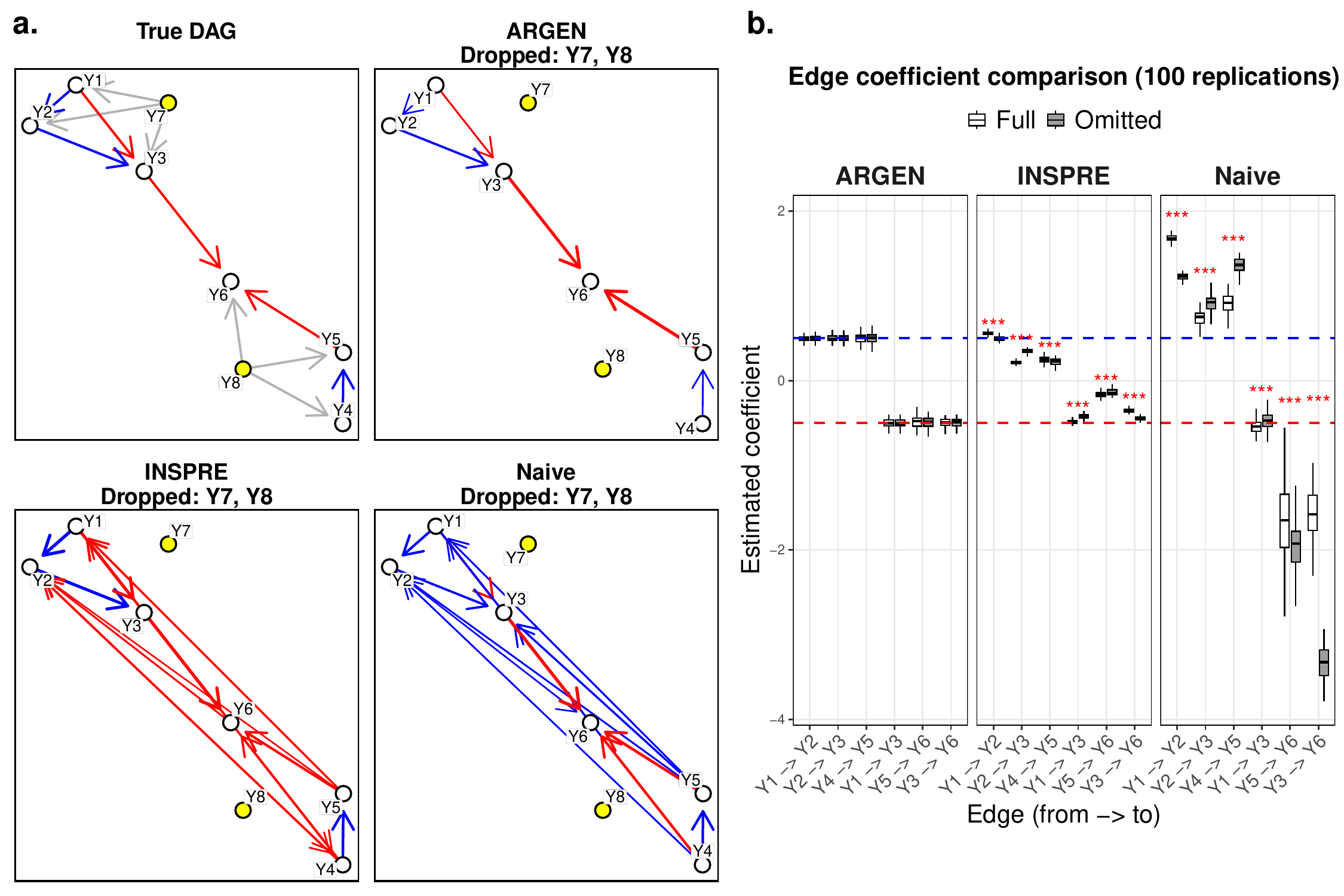}
\caption{\textbf{Simulation experiment results for omitted gene analysis.}\textbf{a.} Visualization of the true data-generating DAG (True DAG) and the DAGs recovered by each of the three methods when genes $Y_7$ and $Y_8$ (yellow nodes) were omitted from the data. The true DAG is defined on the latent true expressions $\mu_j$, but for simplicity the nodes are labeled using the corresponding observed variables $Y_j$. Edges are colored by the sign of the coefficient values (blue: positive; red: negative). Gray edges denote the edges that are omitted in actual applications. \textbf{b.} Boxplots of the estimated coefficients for six edges $\theta_{jk}$ across 100 replicates. Each panel corresponds to one of the three methods, and the blue and red dashed lines indicate the true values 0.5 and -0.5, respectively. Within each panel, the x-axis denotes the six edges, and the boxplot colors distinguish the Full and Omitted analyses. Red stars above each edge indicate the significance of the Wilcoxon rank-sum test comparing the Full and Omitted analyses for that edge.
} \label{fig:Illust_example}
\end{figure}

For a simple illustration, we generated a Perturb-seq dataset ($N=8000, p=8$) involving unmeasured ($U$) and observed ($X$) confounders (\textbf{Fig.}~\ref{fig:Illust_example}a; Sections~\ref{supp-sec:simul_DAG}, \ref{supp-sec:simul_omitgenes} for details and \textbf{Fig.}~\ref{supp-fig:Illust_example_medp},\ref{supp-fig:pn_ratio_contour_omitted} for a larger $p$ scenario). For the experiment, we dropped the confounding genes $Y_7$ and $Y_8$ and applied \methodname{} across 100 replicates. For the case of dropping mediating genes, we refer to Section~\ref{supp-sec:omit_med} and \textbf{Fig.}~\ref{supp-fig:omit_med}. For baseline comparisons, we considered INSPRE \citep{brown2025large} and a Naive GLM approach. INSPRE is the only published method that learns DAG of genes from Perturb-seq data with a global optimization approach. Nonetheless, its causal estimates become vastly susceptible to confounding with even small estimation error on the total cascade effect of a perturbation. The Naive approach uses the causal ordering learned by \methodname{}, but utilizes a naive proxy $\log\eta_{ij}\approx\log(Y_{ij}/l_i+1)$ in the second-stage regression \eqref{eq:second_stage}. Implementation details are in Section~\ref{supp-sec:processing_implementation}.

\methodname{} successfully recovers the underlying subgraph when nodes are omitted, whereas alternative methods fail to (\textbf{Fig.}~\ref{fig:Illust_example}a). Comparing coefficient estimates from analyses utilizing all genes (Full) versus a subset (Omitted), \methodname{} demonstrates both robustness and unbiasedness: its estimates remain consistent across both settings and accurately center on the true values (\textbf{Fig.}~\ref{fig:Illust_example}b). Conversely, competing methods are sensitive to gene omission, exhibiting significant shifts in coefficient estimates (\textbf{Fig.}~\ref{fig:Illust_example}b). Furthermore, these methods yield biased estimates in the `Full' setting, demonstrating vulnerability to unmeasured confounding $U$. These results highlight \methodname{}'s robustness to omitted confounders.

\section{Identification of Causal Essential Gene  Networks from K562 Cells with Perturb-seq Experiments}

As introduced in Section~\ref{sec:motivating_data}, we applied ARGEN to learn causal networks among core essential genes in K562 cells using Perturb-seq data \citep{replogle2022mapping}.
We implemented two sets of applications to properly evaluate and explore the regulatory networks learned from the data. First, for making the evaluation of the causal network feasible from 3D genomics and epigenomics, we initially focused on analyzing the Perturb-seq data in a chromosome specific manner. In other words, we looked at intra-chromosomal regulation of the essential genes (Sections~\ref{sec:direction_robustness} and \ref{sec:epi_3d}). 
This restriction enables us to integrate CRISPRi-derived causal influences with 3D genome architecture and epignetic annotations, providing a natural framework to assess whether inferred causal effects preferentially occur within spatially proximal, epigenetically active neighborhoods. Second, to provide further insights on trans-regulation (inter-chromosomal regulation) among a larger set of essential genes, we analyzed the genes located across chromosomes (Sections~\ref{sec:structure_inter} and \ref{sec:func_epi_eval}). 

\subsection{Empirical evaluation of causal directionality and robustness to omitted variables}\label{sec:direction_robustness}

We first start with analyzing the Perturb-seq data in a chromosome-specific manner and provide evaluations on the causal  network learned from the data.  
We focused on 874 perturbations that had enough number of cells and CRISPRi signal on the targeted genes. For further details on the selection of such genes, refer to Section~\ref{supp-sec:processing_implementation}. For each of the 23 human chromosomes, we collected the genes that reside within the same chromosome and applied \methodname{} to infer a causal gene network. This resulted in  a set of DAGs with around 40 genes and 100 edges per each chromosome on average. 

\begin{figure}[!ht]\centering
   \includegraphics[width=0.8\textwidth]{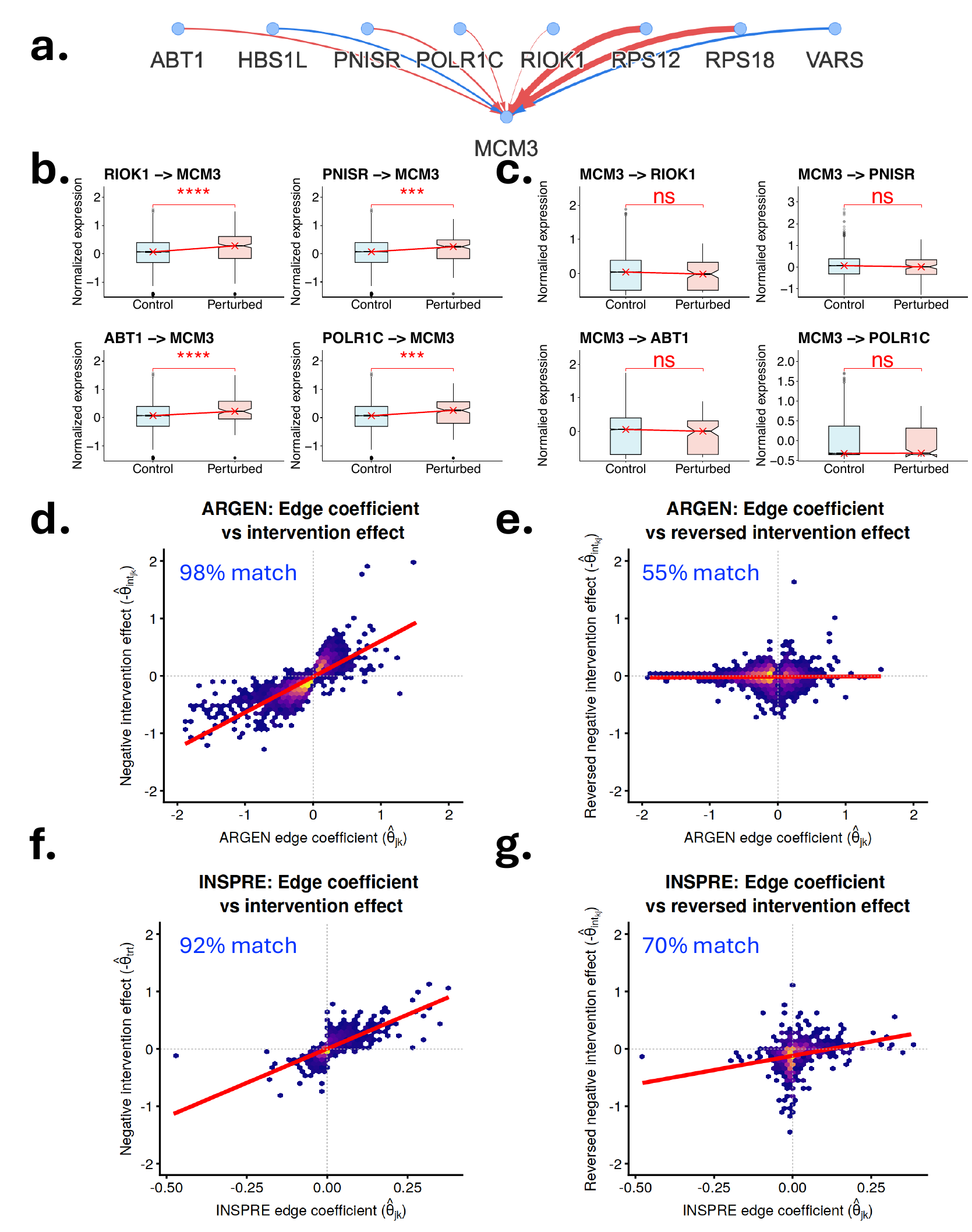}
\caption{\textbf{Directional relationships between genes learned from the intra-chromosomal application of \methodname{}.} \textbf{a.} Visualization of DAG edges from all parent genes to MCM3. Edge colors correspond to the sign of each coefficient (red: negative; blue: positive). \textbf{b.} Comparison of normalized and adjusted MCM3 expression levels between control cells and cells with perturbations across parent genes (RIOK1, PNISR,ABT1 and POLR1C) shown in a. Red stars denote the significance of the Wilcoxon rank-sum test comparing the two groups. \textbf{c.} As in b, but showing the expression levels of each parent gene in control cells and in cells with MCM3 perturbation. The same results for the entire parent genes can be found in \textbf{Fig.}~\ref{supp-fig:mcm3_rest}. \textbf{d.} Comparison between \methodname{} edge z-scores \eqref{edge_zscore} (x-axis) and the intervention effect's z-score, which is multiplied by $-1$ (y-axis) when each parent gene is intervened. All edges across 23 chromosomes are shown. \textbf{e.} As in d, but with interventions applied to the child genes of each target gene rather than to the parent genes. \textbf{f,g.} As in d,e but the edges from INSPRE are displayed.
} \label{fig:coefs}
\end{figure}

We first implemented a sanity check for the directed edges in the DAGs. We checked whether the directional relationships between pairs of genes are indeed causal, instead of being associational. Specifically, as the data is from CRISPRi experiment, where the guide RNA is designed to reduce the expression level of the targeted gene, we would expect to have $\tau_k<0$ in the SEM \eqref{SEM}. Thus, if the estimated edge coefficient $\hat{\theta}_{jk}<0$ for a parent gene $k$ and a child gene $j$, we would expect the expression level of gene $j$ to increase under an intervention on gene $k$; conversely, if $\hat{\theta}_{jk}>0$, we would expect the expression level of gene $j$ to decrease. We 
checked whether such a relationship holds for each target gene. As an illustrative example, we collected the entire parents of the gene MCM3 in chromosome 6 (\textbf{Fig.}~\ref{fig:coefs}a) and checked the relationship. It can be observed from \textbf{Fig.}~\ref{fig:coefs}b and \textbf{Fig.}~\ref{supp-fig:mcm3_rest} that the perturbation on each of the parent genes of MCM3 result in significant increase of MCM3 expression whenever the coefficient $\hat{\theta}_{jk}<0$ and significant decrease of MCM3 expression whenever the $\hat{\theta}_{jk}>0$, aligning with our expectation for CRIPSRi experiments. On the other hand, when MCM3 was perturbed reversely, the expression changes of its parent genes neither exhibit this inverse relationship nor reach comparable levels of significance (\textbf{Fig.}~\ref{fig:coefs}c).

We then checked whether such a result holds for all the edges in the 23 chromosome specific DAGs, and observed that  98\% of the \methodname{} edge coefficients $\hat{\theta}_{jk}$ showed negative association with the intervention treatment effect ($\hat{\theta}_{int_{jk}}$), learned from perturbing a parent gene $k$ of a gene $j$ and evaluating the expression change of the gene (\textbf{Fig.}~\ref{fig:coefs}d). In comparison, only around 55\% of the \methodname{} coefficients showed such relationship with the reversed treatment effect (\textbf{Fig.}~\ref{fig:coefs}e), where a child gene was intervened and the change in expression of a parent node of the child was measured, indicating the sign match of nearly random chance (50\%). Application of INPSRE reveals inferior sign match (92\%) between the edge coefficients and the intervention effects (\textbf{Fig.}~\ref{fig:coefs}f) and higher match (70\%) between the edge coefficients and the reversed intervention effects (\textbf{Fig.}~\ref{fig:coefs}g) than \methodname{}. Application of the Naive GLM method shows similar results. (\textbf{Fig.}~\ref{supp-fig:coef_orient_all}). These results reveal that the directionality of the \methodname{} edges is likely to be causal instead of associational, addressing the causal aspect of our first motivating question (Section~\ref{sec:motivating_data})

\begin{figure}[!ht]\centering
   \includegraphics[width=1\textwidth]{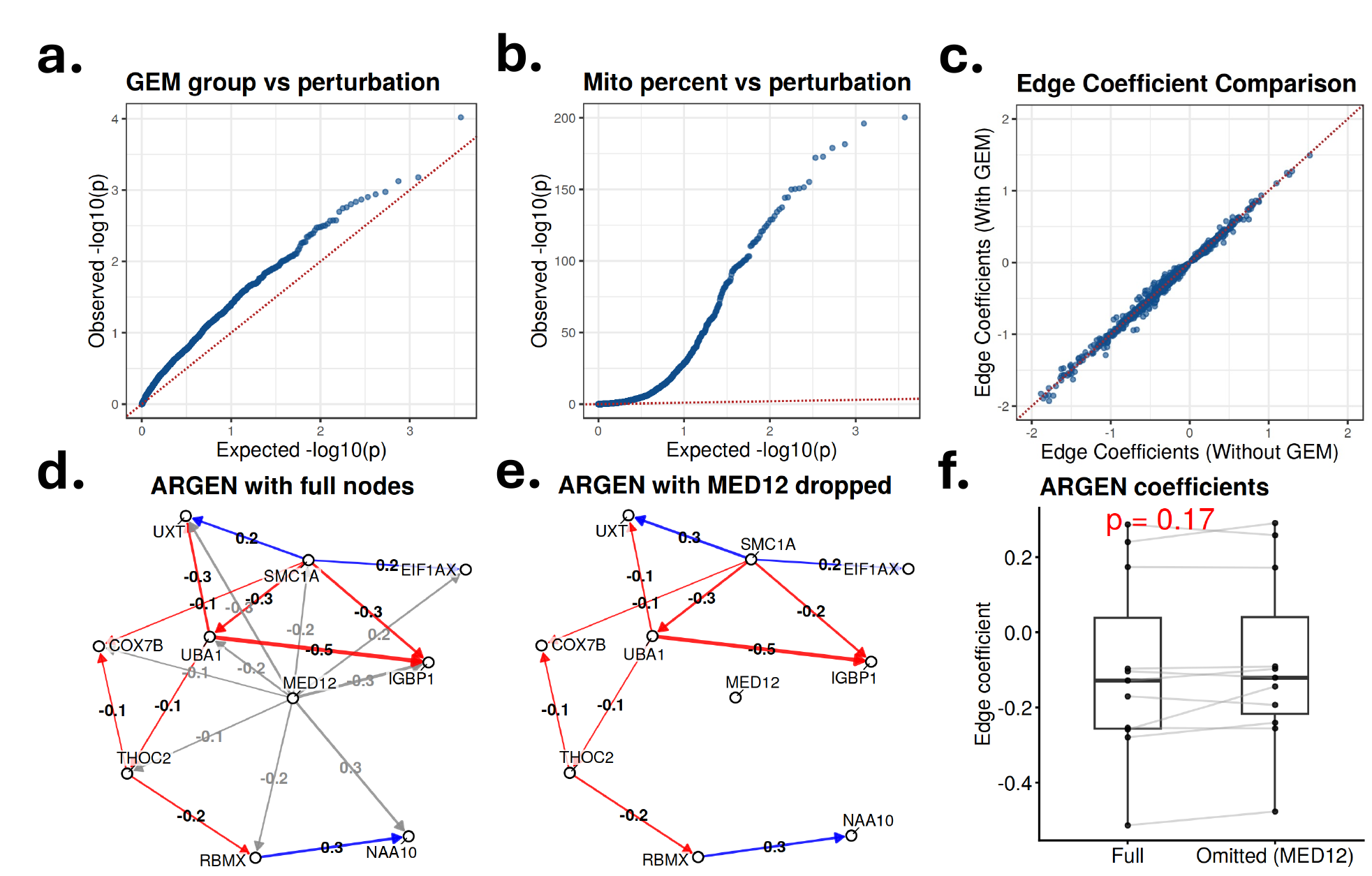}
\caption{\textbf{a.} Quantile-quantile plot of the $-\log_{10}{(\mathrm{p-values})}$ from the ANOVA F-test evaluating the logistic regression of the perturbation indicator on the GEM group variable for each targeted gene in \cite{replogle2022mapping}. \textbf{b.} As in a, but using mitochondrial percentage as the predictor. \textbf{c.} Comparison of \methodname{} edge coefficients estimated from the data with (y-axis) and without (x-axis) the GEM group included as a covariate. \textbf{d.} Subgraph of chromosome X genes confounded by MED12, inferred by \methodname{}. \textbf{e.} As in d, but inferred after omitting the MED12 gene from the analysis. \textbf{f.} Comparison of edge coefficients from the subgraphs inferred with (left) and without (right) MED12 in the analysis.}
\label{fig:sense_confounder}
\end{figure}

It is demonstrated in the Sections~\ref{sec:identify} and \ref{sec:main_simul} that \methodname{} is robust to arbitrary confounders. We conducted additional sensitivity analyses to showcase \methodname{}'s robustness with the motivating data. Note that the GEM (Gel beadin-
EMulsion) group is known to be associating with the gene expression levels of the Perturb-seq data \citep{replogle2022mapping}. On the other hand, our method is robust to omission of such covariate unless it influences the perturbation indicators. Our additional exploratory analyses show that the GEM group does not associate with the perturbation indicators (\textbf{Fig.}~\ref{fig:sense_confounder}a) with no significantly associating ones at $\mathrm{FDR}=0.2$, while the mitochondrial percentage does (\textbf{Fig.}~\ref{fig:sense_confounder}b). 
Therefore, we empirically checked whether the omission of GEM group variable leads to similar causal effect estimates to the case including the GEM group. It turns out that the edge coefficients $\hat{\theta}_{jk}$ from the two cases are almost identical \textbf{Fig.}~\ref{fig:sense_confounder}c (paired t-test $p=0.3$). 

Additionally, we checked whether omission of a confounding gene from the analysis shows similar robustness in coefficient estimation. We compared the graphs learned from two cases, where we include and omit a confounding gene MED12 within chromosome X. The two graphs turn out to be similar whether or not we include MED12 in our analyses (\textbf{Fig.}~\ref{fig:sense_confounder}d,e). A formal analysis also supports that the coefficients do not significantly change ($p=0.17$, \textbf{Fig.}~\ref{fig:sense_confounder}f) based on a paired t-test. These results showcase the robustness of \methodname{} to confounding in real applications, addressing the second key question listed in Section~\ref{sec:motivating_data}.

\subsection{Epigenomic and 3D genomic evaluation of \methodname{} intra-chromosomal causal networks}\label{sec:epi_3d}

Now that we have checked sanity of the causal directions of the \methodname{} edges in the previous section and the robustness to unobserved confounding, we now turn our attention to evaluating the edges with orthogonal biological sources including 3D genomics and epigenomcs. Here, we mainly check whether the external biological evidence support concordance between gene pairs with DAG edges. 

\begin{figure}[!ht]\centering
   \includegraphics[width=0.9\textwidth]{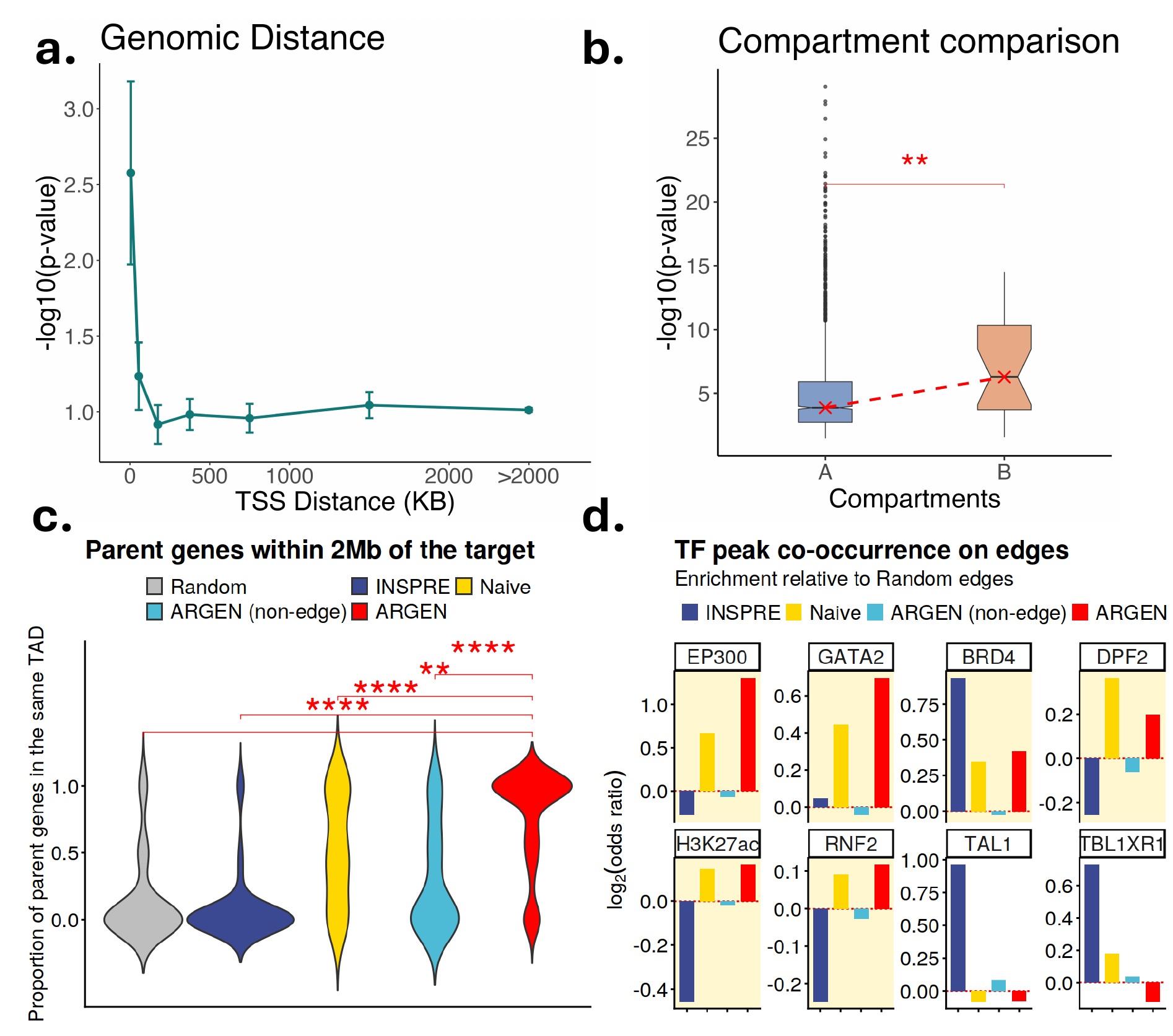}
\caption{
\textbf{Biological evaluation of intra-chromosomal DAG edges based on 3D genomics and epigenomics.} 
\textbf{a.} Relationship between genomic distance in Kb (x-axis) and edge coefficient (log transformed)$p$-values (y-axis). Each error bar represents a mean and $\pm$ SE summary of each of the bins among the following distance intervals: [0–10], (10–150], (150–400], (400–750], (750–1000], (1000–2000], and $>$2000 Kb. \textbf{b.} Box plots comparing the p-values of the edges called as significant by \methodname{}. The edges whose nodes are contained in the A compartment (blue) are compared against those contained in the B compartment (orange).
\textbf{c.} Proportion of parent genes located within the same Topologically Associating Domain (TAD) as the target gene among the parent genes separated from the target gene by less than 2Mb (y-axis). Each data point represents a target gene with at least one parent gene within 2Mb distance from the TSS. The proportion of \methodname{}-significant edges (red) is compared with that of non-significant edges (blue), edges identified by INSPRE (purple), Naive GLM (yellow) and randomly selected edges (gray). 
\textbf{d.} Enrichment of K562-specific transcription factors and the H3K27ac histone mark in promoter regions (1Kb upstream of the transcription start site) of genes involved in each edge. For each protein, we first computed the proportion of random edges with significant peaks at both nodes (genes). We then report $\log_2(\mathrm{odds\ ratio})$ enrichment values relative to random edges for \methodname{}-significant edges (red), non-significant edges (blue), Naive GLM edges (yellow), and INSPRE edges (purple). Each panel corresponds to a protein, and panels with a yellow background highlight proteins for which \methodname{} edges show a higher proportion than random edges.\methodname{} edges exhibit a higher proportion than random edges.
} \label{fig:k562_intra}
\end{figure}

We first investigated the genomic distance pattern of the DAG edges. We specifically gathered the entire $-\log_{10}(\mbox{p-values})$ of $\hat{\theta}_{jk}$ and investigated how the edge signal associates with the genomic distance between parent and child genes. We observe that the genes far apart from the target genes are generally less supported by \methodname{} p-values (\textbf{Fig.}~\ref{fig:k562_intra}a). While genomic distance is known to be an evidence mainly for cis-regulation of genes by their regulatory elements, recent findings \citep{long2022making,liang2025tracing} support functional similarity of genes in a close genomic distance.

We further sought for supports from physical proximity information captured from the same system. We specifically explored the actual physical distance structure of the \methodname{} edges based on an external Hi-C data of K562 cells \citep{rao20143d}. A/B compartments, which form large genomic territories with relatively high (A) or low (B) gene expression compared to other territories, constitute a class of chromatin domain compartmentalization \citep{lieberman2009comprehensive}. Consistent with recent findings \citep{zufferey2021systematic} that co-regulation within chromatin domains is more prevalent among lowly expressed or less efficiently transcribed genes, as well as evidence that closed chromatin regions exhibit elevated co-methylation across loci \citep{fortin2015reconstructing}, we observed that the overall edge signal from \methodname{} was significantly higher in the B compartment than in the A compartment ($p < 5 \times 10^{-3}$). Furthermore, among the gene pairs with Transcription Start Site (TSS) distance less than 2 Mega base pairs (Mb), we checked how many of the parent genes are within the same Topologically Associating Domain (TAD), a contiguous regions of genome within which chromatin interactions are enriched \citep{dixon2012topological}. We investigated the proportion of such regulatory genes in the gene pairs called as significant by \methodname{}'s online FDR procedure in comparison to those in the gene pairs called as not significant. This comparison yielded that those genes within \methodname{} edges show significantly higher proportion ($p<10^{-3}$) than the other set of genes (\textbf{Fig.}~\ref{fig:k562_intra}c). This aligns with recent findings that genes sharing a TAD exhibit elevated regulatory similarity \citep{liang2025tracing}, a phenomenon attributed to TADs acting as physical constraints on phase-separated transcriptional condensates (TC) to amplify the local sharing of regulatory machinery \citep{hnisz2017phase, boija2018transcription}. Furthermore, experimentally disrupting these TCs weakens TAD boundary insulation, confirming their structural-functional interdependence \citep{gamliel2022long}. The \methodname{} edges also showed significantly higher proportion ($p<10^{-3}$) than the ones called by INSPRE and Naive GLM, while the proportion of INSPRE showed no significant difference compared to that of the gene pairs randomly selected (\textbf{Fig.}~\ref{fig:k562_intra}c).

Finally, we checked whether the gene pairs called from \methodname{} showed enriched ChIP-seq signal of binding of a histone mark (H3K27ac) and transcription factors relevant to K562 and studied in \cite{gasperini2019genome}. We quantified the proportion of edges involving enriched ChIP-seq signal in both of the gene promoter regions ($\pm$ 1Kb from TSS), and compared the proportion against that of random gene pairs. We observed that six (\textbf{Fig.}~\ref{fig:k562_intra}d) out of eight ChIP-seq targets showed enrichment in the \methodname{} edges. In comparison, half of the ChIP-seq targets showed lack of enrichment in the INSPRE edges. These results provide the characterization of the \methodname{} edges in terms of 3D genomics and epigenomics, answering the last key question in Section~\ref{sec:motivating_data}.

\subsection{Modular structure of  inter-chromosomal causal gene networks}\label{sec:structure_inter}

\begin{figure}[!ht]\centering
   \includegraphics[width=0.9\textwidth]{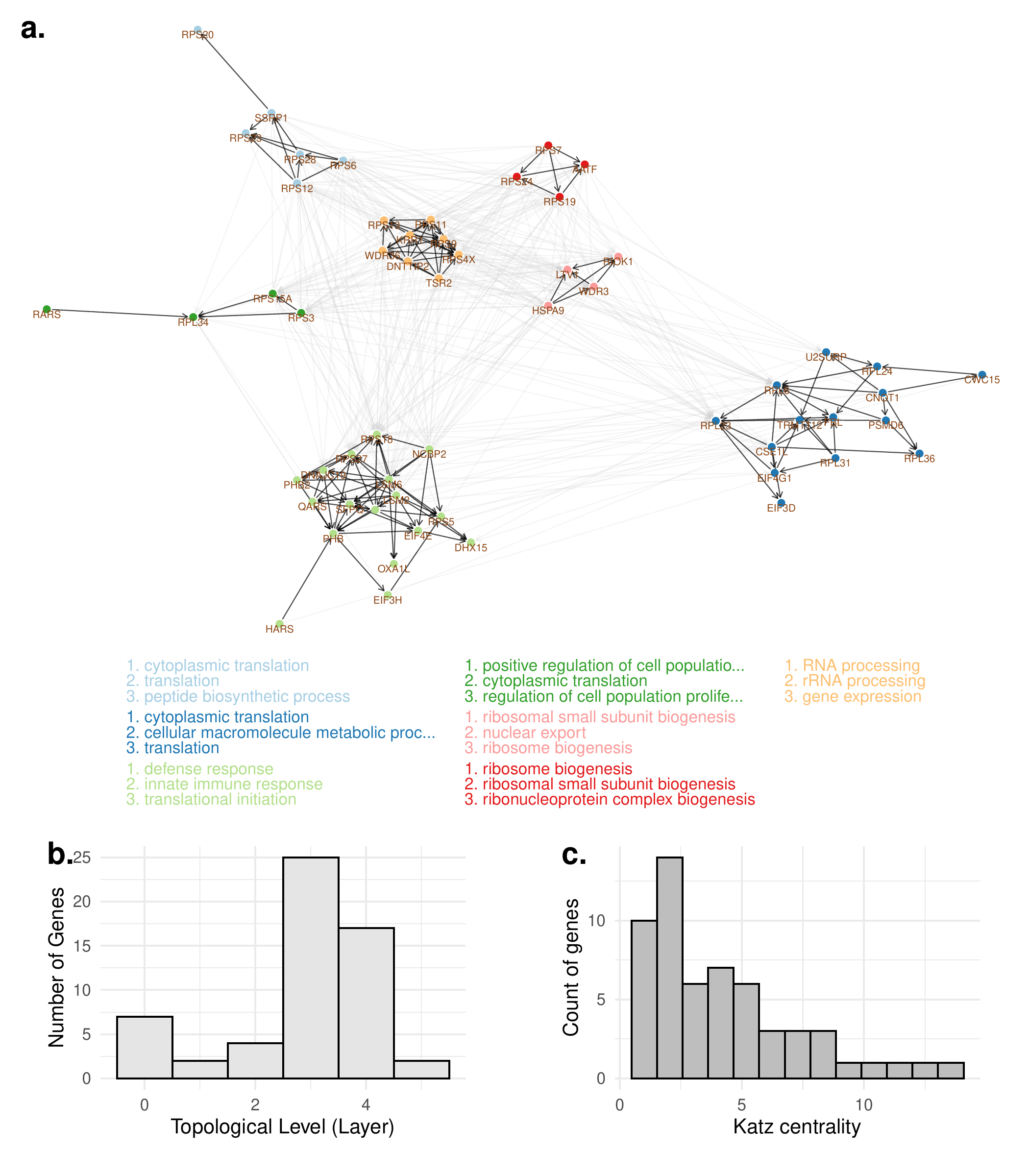}
\caption{\textbf{Global structure of the inter-chromosomal causal gene network of K562 essential genes.} 
\textbf{a.} Visualization of the learned DAG. Nodes are colored according to Louvain module clustering, and for each module, the top three Gene Ontology terms are listed using the corresponding colors. 
\textbf{b.} Histogram of the topological layers of genes in the DAG. 
\textbf{c.} As in b, but showing the distribution of Katz centrality values.
} \label{fig:k562_inter}
\end{figure}

We next applied \methodname{} to learn the causal network of a larger set of the essential 400 genes that were perturbed in at least 100 cells and with significant perturbation effect z-score $-5$ on the target gene (Section~\ref{supp-sec:processing_implementation} for details on gene selection)
These genes play essential roles in  splicing ribosomal, proteasome, cell cycle, transcription, mitochondrial translation and metabolism. 
The model fit resulted in a DAG with 456 edges for $\alpha=0.01$ for both the descendant search and parent search (\textbf{Fig.}~\ref{fig:k562_inter}a).

We examined the structure of the inferred inter-chromosomal network by applying Louvain clustering with edge z-scores \eqref{edge_zscore} as weights, yielding 7 modules annotated by their top three Gene Ontology (GO) terms (\textbf{Fig.}~\ref{fig:k562_inter}a). The clusters form distinct and also coherent functional modules dominated by core cellular processes, consistent with essential-gene categories in \cite{replogle2022mapping}. Notably: (i) \textit{cytoplasmic translation}, represented by two (blue) modules enriched for ribosomal proteins (RPL/RPS), reflecting the translational demands of proliferative K562 cells \citep{khajuria2018ribosome};
(ii) \textit{defense response}, represented by a (light green) module enriched for innate antiviral factors (PHB2/EIF4E/SFPQ), which couple mitochondrial signaling, translational control, and transcriptional derepression to the rapid production of type I interferons \citep{refolo2020mitochondrial, herdy2012translational, imamura2014long}; and (iii) \textit{RNA processing}, represented by a (orange) module enriched for nucleolar processing factors (WDR36/KRR1), which couple pre-rRNA cleavage and transcript maturation to the rapid proliferation of K562 cells \citep{skarie2008primary, singh2021nucleolar}.

Although we focus here on an inter-chromosomal causal influence network, such long-range structure is biologically expected rather than surprising. Core essential processes, such as cytoplasmic translation, defense response, and RNA processoing, are encoded by genes distributed across many chromosomes and are regulated by trans-acting factors whose targets span the genome. The fact that our inferred inter-chromosomal network clusters into functionally coherent modules  therefore suggests that the model is capturing biologically meaningful regulatory programs rather than merely reflecting genomic proximity.

The topological structures learned by \methodname{} provide compelling structural insights. The hierarchical structure of the DAG in \textbf{Fig.}~\ref{fig:k562_inter}b reveals that the learned architecture is shallow (average depth: 2.8). This suggests that most genes function as near-direct regulators, with regulatory signals rarely propagating through long multi-step chains. This is consistent with previous findings that essential cellular processes are organized into compact regulatory modules rather than extensive cascading effects \citep{alon2007network}. In addition, a summary of the DAG with Katz centrality indicates that there are a few hub nodes with high centrality (\textbf{Fig.}~\ref{fig:k562_inter}c). It is widely observed that gene networks are primarily scale-free and possess highly right-skewed degree distributions \citep{barabasi1999emergence,hecker2009gene}. Examining the in- and out-degree distributions separately (\textbf{Fig.}~\ref{supp-fig:degree_comparison}a,b) reveals a similar right-skewed pattern for \methodname{}. In contrast, the out-degree of INSPRE (\textbf{Fig.}~\ref{supp-fig:degree_comparison}c) and the in-degree of the Naive GLM (\textbf{Fig.}~\ref{supp-fig:degree_comparison}f) exhibit bell-shaped distributions, mismatching with these prior observations. Together, these structural features directly address our first core question (Section~\ref{sec:motivating_data}) regarding the causal architecture of K562 essential processes.

\subsection{Functional characterization and epigenetic evaluation of the inferred gene network}\label{sec:func_epi_eval}

\begin{figure}[!ht]\centering
   \includegraphics[height=0.56\textheight,width=0.9\textwidth]{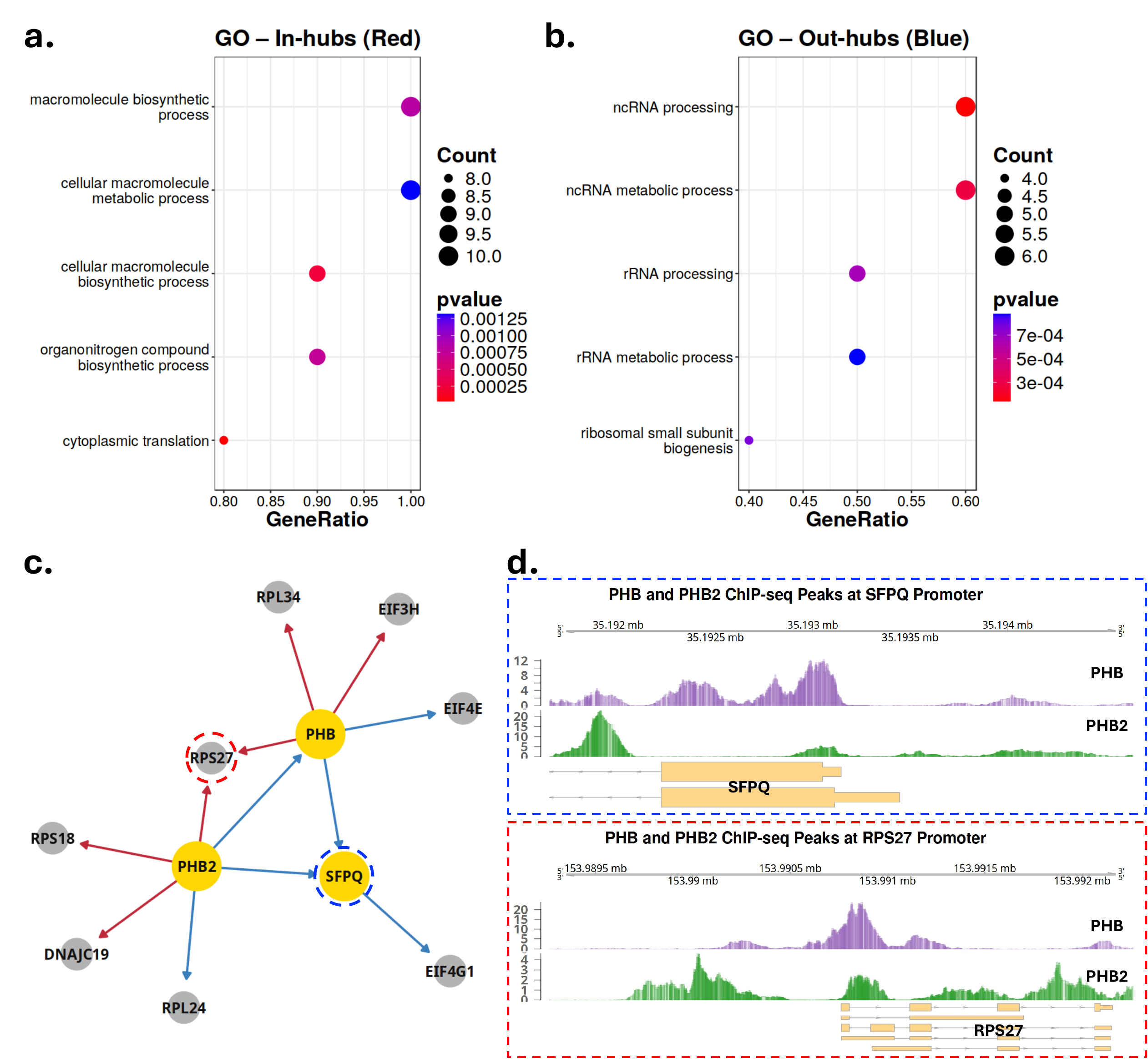}
\caption{\textbf{Functional and epigenetic characterization of \methodname{} gene network.} \textbf{a,b.} Top five Gene Ontology terms for in-hubs (a) and out-hubs (b). \textbf{c.} Sub-graph of \methodname{} DAG extracting children genes of PHB, PHB2 and SFPQ. Here, yellow nodes indicate that the gene is a TF, whose Chip-seq is sequenced and listed in ENCODE \citep{encode2012integrated}. \textbf{d.} Top: $\pm1.5$ Kb promoter region of the SFPQ gene TSS and the ChIP-seq signals of PHB (purple) and PHB2 (green). Signal tracks display the $-\log_{10}(\mbox{p-value})$ generated via MACS2 \citep{zhang2008model}. Bottom: Same as the top panel but the promoter region of RPS27 gene was displayed.
} \label{fig:k562_function}
\end{figure}

For a functional characterization of the genes with high connectivity, we focused on 10 in-hub and 10 out-hub genes, defined as those with the largest in- and out-degree, respectively (\textbf{Fig.}~\ref{supp-fig:in_out}). GO analysis for each group revealed distinct functionalities: In-hubs were enriched for cytoplasmic translation and cellular amide metabolic process, whereas Out-hubs were enriched for ribosome biogenesis and RNA processing. This division reflects established biology because translation-related genes are highly dosage-sensitive and tightly regulated \citep{torres2007effects}, while ribosome biogenesis and RNA-processing genes function as upstream controllers of growth and transcriptional output \citep{lempiainen2009growth}. In comparison, INSPRE in-hub genes showed enrichment in ribosome biogenesis and RNA processing (\textbf{Fig.}~\ref{supp-fig:enrich_comparison}c) and the out-hub genes from Naive GLM showed high enrichment in cytoplasmic translation (\textbf{Fig.}~\ref{supp-fig:enrich_comparison}f), misaligning with what is supported by the literature.

Lastly, we evaluated the validity of gene regulatory relationship learned from \methodname{}. We specifically focused on the transcription factors PHB, PHB2 and SFPQ, which had non-zero degree in the graph (\textbf{Fig.}~\ref{fig:k562_function}c). 

First, it turned out that RPS27 and SFPQ, co-regulated by PHB and PHB2, show enriched Chip-seq signals of PHB and PHB2 on their promoter regions (\textbf{Fig.}~\ref{fig:k562_function}d). Similarly, each child gene of the TF genes shows enriched TF Chip-seq signal on its promoter (\textbf{Fig.}~\ref{supp-fig:all_tf}). Within the sub-network involving the TFs, the paths linking PHB to EIF family genes, such as EIF4E and EIF4G1, are corroborated by recent evidence demonstrating a direct physical interaction between prohibitins and the eIF4F translation initiation complex \citep{largeot2023inhibition}. The regulation of DNAJC19 by PHB2 is supported by proteomic analyses identifying DNAJC19 as a direct binding partner within the mitochondrial PHB2 interactome \citep{richter2014dnajc19}. Furthermore, the influence of PHB2 on ribosomal genes (RPS27, RPL24) aligns with previous finding that these proteins co-assemble into broader macromolecular complexes \citep{mugabo2018elucidation}.
These results add biological fidelity of key regulatory features within the causal gene network.

\section{Discussion}
In this work, we developed a new methodological framework for learning causal gene networks from Perturb-seq experiments in the presence of unobserved confounding. Although CRISPR-based perturbation provides a powerful source of exogenous variation, latent biological and technical factors remain pervasive in single-cell data and can distort standard causal estimates. By combining ideas from proxy adjustment and instrumental-variable reasoning, our approach leverages the multi-perturbation structure of Perturb-seq to produce unbiased causal effect estimates even when confounders are arbitrary and unmodeled. Through both simulations and real-data applications, we demonstrated that the method improves substantially upon existing approaches that do not explicitly account for latent structure, yielding more accurate reconstruction of the underlying DAG governing gene expression.

Our application to a large-scale CRISPRi Perturb-seq experiment in K562 cells highlights the biological insights that can be gained from robust causal DAG inference. At the genome-wide scale, the inferred inter-chromosomal network clustered into coherent modules reflecting core essential processes, including cytoplasmic translation, RNA processing, and defense response. These findings are consistent with known functional dependencies among essential genes and illustrate that causal relationships inferred from Perturb-seq reflect functional organization rather than mere genomic proximity.

Several extensions merit consideration. First, our model assumes perturbation effects propagate through steady-state transcriptional responses, whereas some regulatory interactions may operate at translational timescales. Incorporating time-resolved single-cell perturbation data may improve resolution of dynamic regulatory mechanisms. Second, our framework is designed to provide an accurate causal network over the genes intervened in Perturb-seq experiments. While this is practically meaningful given that perturbed genes are typically chosen for biological relevance, extending the framework to partial intervention settings and characterizing the identifiable DAGs remain important. Finally, although we analyzed a well-characterized essential-gene CRISPRi screen, extending the method to more context-specific regulatory settings such as developmental systems or tumor microenvironments will require further methodological refinements and experimental validation.

\if1\anon
{
\section*{Acknowledgements} This research is supported by NIH grants GM129781 and HG013841.

\section*{Data Availability Statement}
Data and R codes to reproduce the results are available in the \href{https://github.com/kmp0223/ARGEN}{ARGEN Github repository}.
} \fi

\bibliography{bibtex}

\end{document}